\documentclass[iop]{emulateapj}

\usepackage{epsfig}

\def\s{\phantom}

\def\amin{\ifmmode^{\prime}\else$^{\prime}$\fi}
\def\asec{\ifmmode^{\prime\prime}\else$^{\prime\prime}$\fi}

\def\simgt{\lower.5ex\hbox{$\; \buildrel > \over \sim \;$}}
\def\simlt{\lower.5ex\hbox{$\; \buildrel < \over \sim \;$}}

\newcommand\chandra{{\it Chandra}}
\newcommand\xmm{{\it XMM-Newton}}

\newcommand\fermi{{\it Fermi}}
\newcommand\swift{{\it Swift\/}}
\newcommand\nustar{{\it NuSTAR}}

\newcommand\hawcsrc{{2HWC~J1928$+$177}}
\newcommand\xraysrc{{CXO1928}}

\newcommand\pulsar{{PSR~J1928$+$1746}}

\newcommand\eflux{{erg\,cm$^{-2}$\,s$^{-1}$}}

\shorttitle{Multi-wavelength observations of \hawcsrc}
\shortauthors{Mori et al.}

\begin{document}

\title{Multi-wavelength  observations of 2HWC~J1928+177: dark accelerator or new TeV gamma-ray binary?} 















\author{Kaya Mori\altaffilmark{1}, Hongjun An\altaffilmark{2}, Qi Feng\altaffilmark{3}, Kelly Malone\altaffilmark{4}, Raul R. Prado\altaffilmark{5}, Yve~E.~Schutt\altaffilmark{1}, Brenda~L.~Dingus\altaffilmark{4}, E.V.~Gotthelf\altaffilmark{1},   Charles~J.~Hailey\altaffilmark{1}, Jeremy Hare\altaffilmark{6}, Oleg Kargaltsev\altaffilmark{7}, Reshmi Mukherjee\altaffilmark{8}}

\altaffiltext{1}{Columbia Astrophysics Laboratory, Columbia University, New York, NY 10027, USA}

\altaffiltext{2}{Department of Astronomy and Space Science, Chungbuk National University, Cheongju 28644, Republic of Korea}

\altaffiltext{3}{Physics Department, Columbia University, New York, NY 10027, USA}

\altaffiltext{4}{Los Alamos National Laboratory, Los Alamos, NM, USA}

\altaffiltext{5}{DESY, Platanenallee 6, 15738 Zeuthen, Germany}

\altaffiltext{6}{NASA Postdoctoral Program Fellow. NASA Goddard Space Flight Center, Greenbelt, MD 20771, USA}

\altaffiltext{7}{Department of Physics, The George Washington University, 725 21st Street NW, Washington, DC 20052, USA}

\altaffiltext{8}{Department of Physics and Astronomy, Barnard College, Columbia University, NY 10027, USA}

\email{kaya@astro.columbia.edu}

\begin{abstract}

\hawcsrc\ is a Galactic TeV gamma-ray source detected by the High Altitude Water Cherenkov (HAWC) Observatory  up to $\sim56$~TeV. The HAWC source, later confirmed by H.E.S.S., still remains unidentified as a dark accelerator since there is no apparent supernova remnant or pulsar wind nebula detected in the lower energy bands. The radio pulsar \pulsar, coinciding with the HAWC source position, has no X-ray counterpart. 
Our SED modeling shows that inverse Compton scattering in the putative pulsar wind nebula can account for the TeV emission only if the unseen nebula is extended beyond $r\sim4$\amin. Alternatively, TeV gamma rays may be produced by hadronic interactions between relativistic protons from an undetected supernova remnant associated with the radio pulsar and a nearby molecular cloud G52.9$+$0.1. 
\nustar\ and \chandra\ observations detected a variable X-ray point source within the HAWC error circle, potentially associated with a bright IR source.
The X-ray spectra can be fitted with an absorbed power-law model with $N_{\rm H} = (9\pm3)\times10^{22}$~cm$^{-2}$ and $\Gamma_X = 1.6\pm0.3$ and exhibit long-term X-ray flux variability over the last decade. If the X-ray source, possibly associated with the IR source (likely an O star), is the counterpart of the HAWC source, it may be a new TeV gamma-ray binary powered by collisions between the pulsar wind and stellar wind.  
Follow-up X-ray observations are warranted to search for diffuse X-ray emission and determine the nature of the HAWC source. 

\end{abstract}



\section{Introduction} \label{sec:intro}

Over the last two decades, the advent of ground-based imaging air Cherenkov telescopes (IACTs) such as H.E.S.S., VERITAS and MAGIC uncovered  a large number of TeV gamma-ray sources, most of which are associated 
with either pulsar wind nebulae (PWNe) or supernova remnants (SNRs). Identifying the nature of Galactic TeV gamma-ray sources is crucial for understanding the cosmic-ray acceleration mechanisms up to the TeV or PeV energy bands. Several TeV observations suggested  the existence of the most extreme cosmic particle accelerator, the so-called {\it Pevatron}, in the Galactic Center \citep{Hess2016} or elsewhere in our Galaxy \citep{Xin2019}. 
More recently, the High Altitude Water Cherenkov (HAWC) Observatory opened a new window for probing gamma-ray sources at even higher energies than the IACTs,  $>$100 TeV  \citep{2017ApJ...843...39A,2019ApJ...881..134A}. HAWC is a TeV gamma-ray telescope equipped with 300 water Cherenkov detectors (WCDs) which directly detects air shower particles produced by TeV gamma rays in the upper atmosphere and can collect data for sources continuously under all  weather conditions. Therefore, the HAWC Observatory is more sensitive than IACTs at energies above $\sim10$~TeV and uniquely explores astrophysical sources in the highest energy gamma-ray band up to a few hundred TeV. 
About a half of the 39 Galactic HAWC sources have not been associated with previously known TeV sources detected by IACTs \citep{Abeysekara2017}. 

\begin{figure}[h]      
\begin{center} 
  \includegraphics[width=1.1\linewidth]{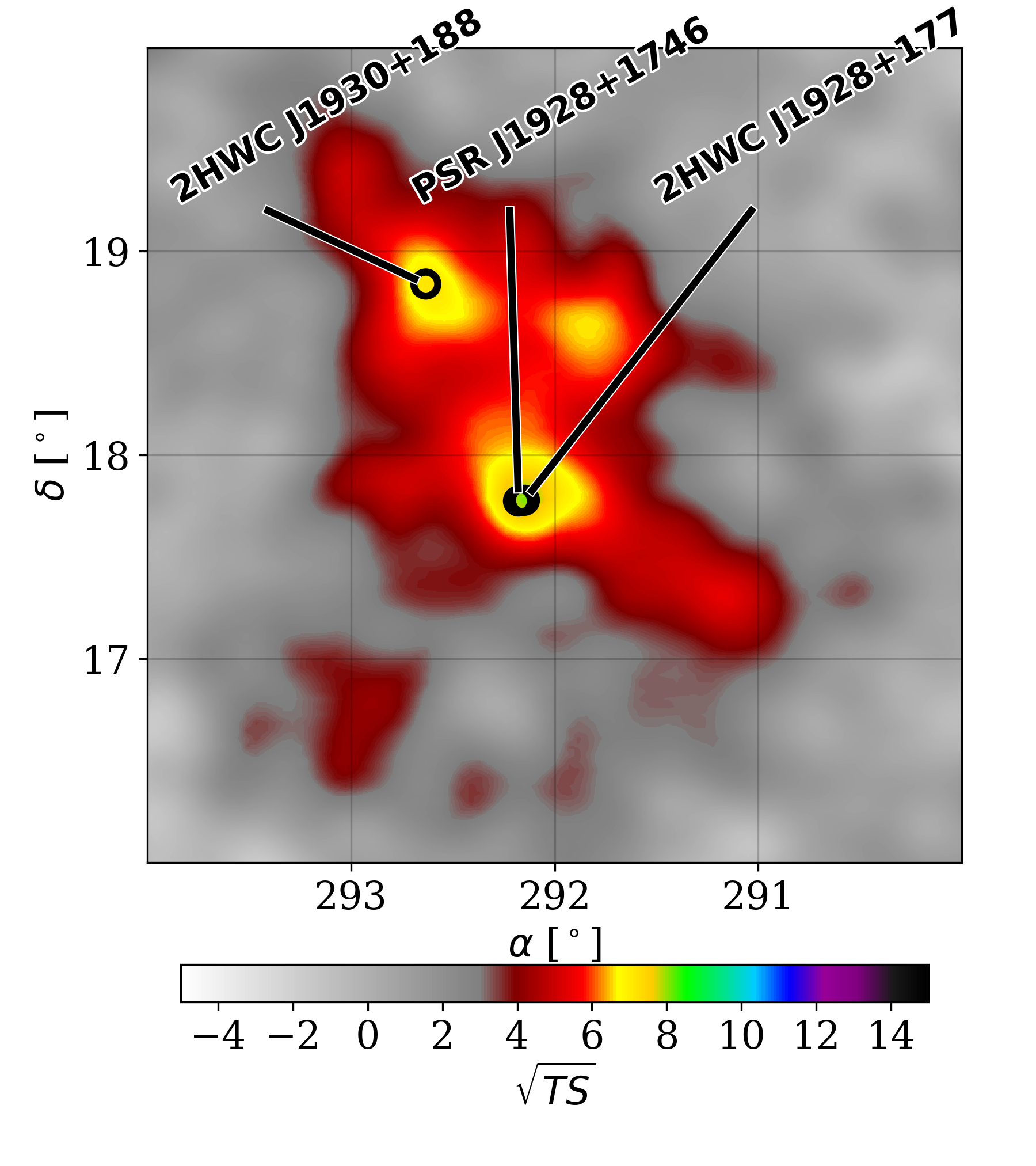}
  \caption{HAWC TeV skymap of the region around \hawcsrc\ and 2HWC J1930$+$188, both of which are modeled as point sources. The radio pulsar \pulsar\ overlaps with the position of \hawcsrc. The figure has been adapted from \cite{Abeysekara2017}. }
\label{fig:hawc_image}
\end{center} 

\end{figure}

\hawcsrc\ is one of the Galactic TeV sources detected by the HAWC standard point source search \citep[][Figure~\ref{fig:hawc_image}]{Abeysekara2017}. Assuming a single power-law spectrum ($N(E) \propto E^{-\Gamma}$) in the TeV band, the best-fit photon index is inferred to be $\Gamma = 2.56 \pm 0.14$. The centroid of the source is 0.03$^\circ$ away from the radio pulsar PSR J1928$+$1746 and $\sim$1.18$^{\circ}$ away from another HAWC source, 2HWC J1930$+$188, which is associated with the SNR G54.1$+$0.3. 
The measured TeV flux in the whole region (Figure~\ref{fig:hawc_image}) for an extended source hypothesis is significantly larger than the sum of the point source fluxes for \hawcsrc\ and 2HWC J1930$+$188, which may imply that one or both of the sources are actually extended. 
{ H.E.S.S. later confirmed the detection of \hawcsrc\ in the  TeV band after applying a different background subtraction method such that the H.E.S.S. maps are more comparable with the HAWC skymaps \citep{JB2019}}. 

Taking into account the luminosity and spectral index of this source in the TeV band, the emission seems to match what is expected for a TeV PWN~\citep{Hess2018}. 
The 83-kyr-old radio pulsar PSR~J1928$+$1746 \citep{Cordes2006} is a good counterpart candidate, as its position coincides well with the HAWC source position (Figure~\ref{fig:hawc_image}). 
However, the pulsar shows no PWN in the radio and X-ray bands despite its relatively high spin-down power ($\dot{E} = 1.6\times10^{36}$~erg\,s$^{-1}$).  There are two nearby \fermi\ sources in the 4FGL catalog \citep{Fermi2019}, but their positions do not overlap with the HAWC source; therefore, their association with the TeV emission is unlikely. 
 For a SNR explanation to work, more than 10$\%$ of the SN energy would need to go towards the acceleration of $>$ 1 TeV protons, which is unlikely~\citep{Lopez-Coto2017}.  
Given the lack of apparent environments for TeV photon production, 
\hawcsrc\ is possibly a unique TeV gamma-ray source whose origin is currently uncertain.

In this paper, we present  multi-wavelength observations of \hawcsrc\ and our investigation of the nature of the TeV emission. We first review recent gamma-ray  observations of the HAWC source by VERITAS and H.E.S.S. (\S\ref{sec:tev}) and by \fermi-LAT (\S\ref{sec:gev}). We then describe the X-ray observations of the field made with \nustar\ and \chandra\, and present the analysis results (\S\ref{sec:obs}). Then, in \S\ref{sec:discussion}, we discuss the nature of \hawcsrc\ using multi-wavelength spectral energy distribution (SED) data and models representing three possible scenarios: a pure leptonic case (PWN), a hadronic accelerator model (dark accelerator), and a TeV gamma-ray binary  scenario. Finally, we summarize our results and future prospects in \S\ref{sec:summary}.

\section{TeV gamma-ray observations of \hawcsrc\ with VERITAS and H.E.S.S} \label{sec:tev} 


VERITAS had previously observed the pulsar \pulsar\ but only detected a 1.2-$\sigma$ excess at the pulsar location~\citep{Acciari2010}. { The non-detection by VERITAS seems to imply that there may be extended emission that is larger than VERITAS’s point spread function (PSF). }
VERITAS recently published a study of 14 HAWC sources in the 2HWC catalog that are not associated with previously known TeV sources~\citep{Abeysekara2018}. One of the regions discussed in detail was the part of the sky containing 2HWC J1930$+$188 and \hawcsrc. 2HWC J1930$+$188 is associated with the TeV source previously identified by VERITAS, VER J1930$+$188~\citep{Acciari2010}, a known TeV PWN G54.1$+$0.3. Further analysis with VERITAS data found no emission from \hawcsrc\ in either a point source (angular extension radius $<0.1^\circ$) or an extended source (angular extension radius $<0.23^\circ$) search. The flux upper limits (99\% confidence level) derived by VERITAS for \hawcsrc\ are $< 6.8\times 10^{-13}$~\eflux and $< 2.2 \times 10^{-12}$~\eflux, for the point and an extended source search, respectively. 
The total exposure time for this analysis is 44 hours (archival data analyzed for VERITAS ranges from 2007 to 2015) and the upper limits are calculated 
above an energy threshold of 460 GeV, assuming the photon index measured by HAWC ($\Gamma = 2.56$). 

{ \hawcsrc\ was not previously detected in the original analysis of the region observed as part of the H.E.S.S. Galactic Plane Survey \citep{Hess2018} using nearly 2700 hours of quality-selected data. H.E.S.S. detected SNR~G54.1$+$0.3 (HESS J1930$+$188) but did not report any very high energy gamma-ray emission coincident with the direction of \hawcsrc. 
A H.E.S.S. study was recently carried out for comparing the Galactic plane as seen by HAWC and H.E.S.S.~\citep{JB2019}. Using a different background estimation for the H.E.S.S. analysis than employed previously, \cite{JB2019} reported a detection of \hawcsrc\ as well as two other point sources having detection significances above $5\sigma$, each of which is less than half a degree away from the corresponding HAWC counterpart. It is interesting to note that this new technique seems promising for comparing IACT data with HAWC detections. These VERITAS and H.E.S.S observations imply that the HAWC source may be extended up to $\Delta\theta\sim0.4^\circ$. Alternatively, the source may be variable as another TeV gamma-ray binary HESS J0632$+$057 was not initially detected by VERITAS \citep{Acciari2009}.}

\section{\fermi-LAT observations and data analysis} \label{sec:gev} 

{ In the fourth  \fermi\ LAT catalog \citep{Abdollahi2020}, there are two GeV sources (4FGL J1928.4$+$1801 and 4FGL J1929.0$+$1729) within about 0.3$^\circ$ from the source position of 2HWC J1928$+$177. However, the flux  extrapolated from their GeV spectra lies far below the HAWC source flux in the TeV band. It is thus unlikely that the two 4FGL sources are associated with the HAWC source.}  
We searched for a point source at the position of \hawcsrc\ in
\fermi-LAT data from the start of the mission to 2020 February. 
To avoid source confusion, we select photons between 1 GeV and 2 TeV, as the PSF of LAT improves with energy and the 68\% containment radius
is below 1$^\circ$ above 1 GeV. A power-law model leaving photon index 
and normalization free at the position of \hawcsrc, and all other
sources in the 4FGL catalog were included in a likelihood analysis. 
No significant detection of \hawcsrc\ was found, with a test
statistic value of 4.1 (~2 $\sigma$), and a 95\% upper limit on the energy flux above 1 GeV of $2.8\times10^{-12}\; \text{erg}\;\text{cm}^{-2}\; \text{s}^{-1}$. { Assuming an extended source with $\Delta\theta=0.4^\circ$ did not change the GeV flux limits significantly ($\simlt 10$\%). } The \fermi-LAT flux upper limits are used for multi-wavelength SED fitting in \S\ref{sec:discussion}.

\section{X-ray observations and data analysis} \label{sec:obs}

We used a 90-ks \nustar\ observation (ObsID 30362002002) taken in 2017 June as a part of the \nustar-VERITAS-HAWC Legacy program which includes observations of PWN DA~495 \citep{Coerver2019} and the  TeV gamma-ray binary HESS~J0632$+$057 \citep{Archer2019}, and two archival \chandra\ observations (ObsIDs 9081 and 22145) which were taken with 10~ks exposures in 2008 and 2019, respectively. We also considered a handful of \swift/XRT observations of the field, but their short exposures yielded only a few counts in each observation, and so the \swift\ data are not very useful even after combining all the data. 

\subsection{Data reduction} 

The \nustar\ data were processed and analyzed using the {\tt NUSTARDAS} v1.7.1 integrated in the HEASOFT 6.25 software package along with the \nustar\ Calibration Database (CALDB) files v20190513. The \chandra\ data were reprocessed with the {\tt chandra\_repro} tool of CIAO 4.11 to use the most recent calibration database. 

\subsection{Image analysis} 
We first obtained the \nustar\ background of the field for an image analysis. We used the {\tt nuskybgd} software \citep{Wik2014} to model the spatial and energy-dependent cosmic X-rays and a detector background. The background model components were determined by fitting the observed spectra in several source-free regions. The background spectra showed no significant Fe line at $E\sim6-7$~keV, indicating that the contamination from the Galactic ridge X-ray emission is negligible \citep{Mori2015}. 

\begin{figure}
\begin{center} 
  \includegraphics[width=0.9\linewidth, angle=90]{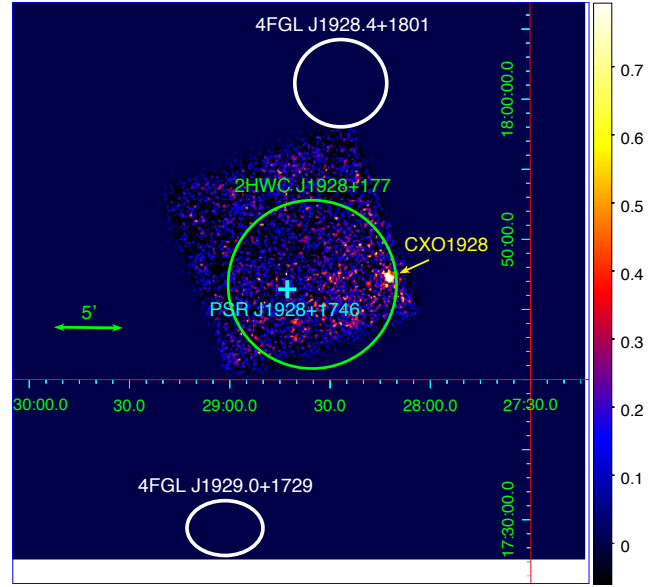}
  \caption{\nustar\ 3--20~keV background-subtracted image of \hawcsrc\ exhibiting the only  X-ray source (\xraysrc) in the \nustar\ FOV. The green circle indicates the position of the HAWC source and its $1\sigma$ error. The positions of two  \fermi\ sources with their 95\% C.L. errors are shown as the white ellipses, while the cyan cross corresponds to the radio pulsar PSR~J1928$+$1746. }
  \label{fig:nustar_image}
\end{center}
\end{figure}

Figure~\ref{fig:nustar_image} shows the  background-subtracted \nustar\ image, after combining the two detector modules, in the 3--20~keV energy band. In the $13\amin\times13\amin$ FOV, \nustar\ detected 
one hard X-ray source which coincides with a point source, CXO~J192812.0$+$174712 (\xraysrc\ hereafter), originally detected by \chandra. \xraysrc\ is the brightest X-ray source detected in the hard X-ray band up to $\sim20$~keV within the HAWC source position error circle. Using the CIAO tool {\tt wavdetect} and the 2019 \chandra\ observation data, where \xraysrc\ was observed near the on-axis position, we determined the \chandra\ position to be RA = 19:28:12\asec.05 and DEC = 17:47:13\asec.35 (J2000) with the 1-$\sigma$ statistical error of 0.9\asec.

We detected no X-ray source at the position of the radio pulsar, confirming the non-detection by \chandra. Using the 20~ksec \chandra\ ACIS data, we determined the 2--8~keV flux upper limits (90\%) of several circular regions around the radio pulsar position. For each source region, we computed the Poisson probability of detecting source counts over background counts (which were obtained from a region elsewhere on the ACIS FOV and corrected for different aperture sizes) and assumed a power-law spectrum with $\Gamma = 2$ for calculating X-ray fluxes. We found that a 90\% upper limit flux (unabsorbed) in  the 2--8~keV band  is $7\times10^{-15}, 7\times10^{-14}$ and $3\times10^{-13}$~\eflux\  for $r=3$\asec, 1\amin\ and 4\amin, respectively. These \chandra\ flux limits 
are useful for constraining diffuse X-ray emission in both the leptonic and hadronic scenarios discussed in \S\ref{sec:pulsar} and \S\ref{sec:hadronic}, respectively.

\subsection{Spectral and timing analysis} \label{sec:analysis}

\begin{figure}[ht]
\begin{center}
\includegraphics[angle=-90, width=1.1\linewidth]{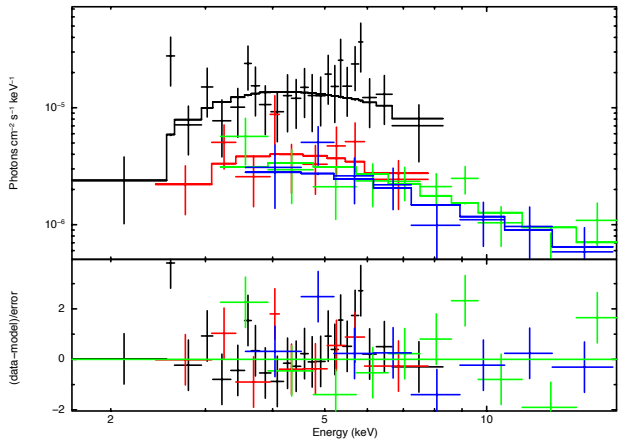}
\caption{\chandra\ and \nustar\ spectra jointly fit with an absorbed power-law model. To account for X-ray
 flux variation, we fit a flux
normalization factor between the \chandra\ ACIS (black: 2008, red: 2019) and the \nustar\ spectra (green: FPMA, blue: FPMB).\label{fig:spectra}
}
\end{center} 
\end{figure}
\begin{figure}
\begin{center} 
  \includegraphics[width=1.0\linewidth]{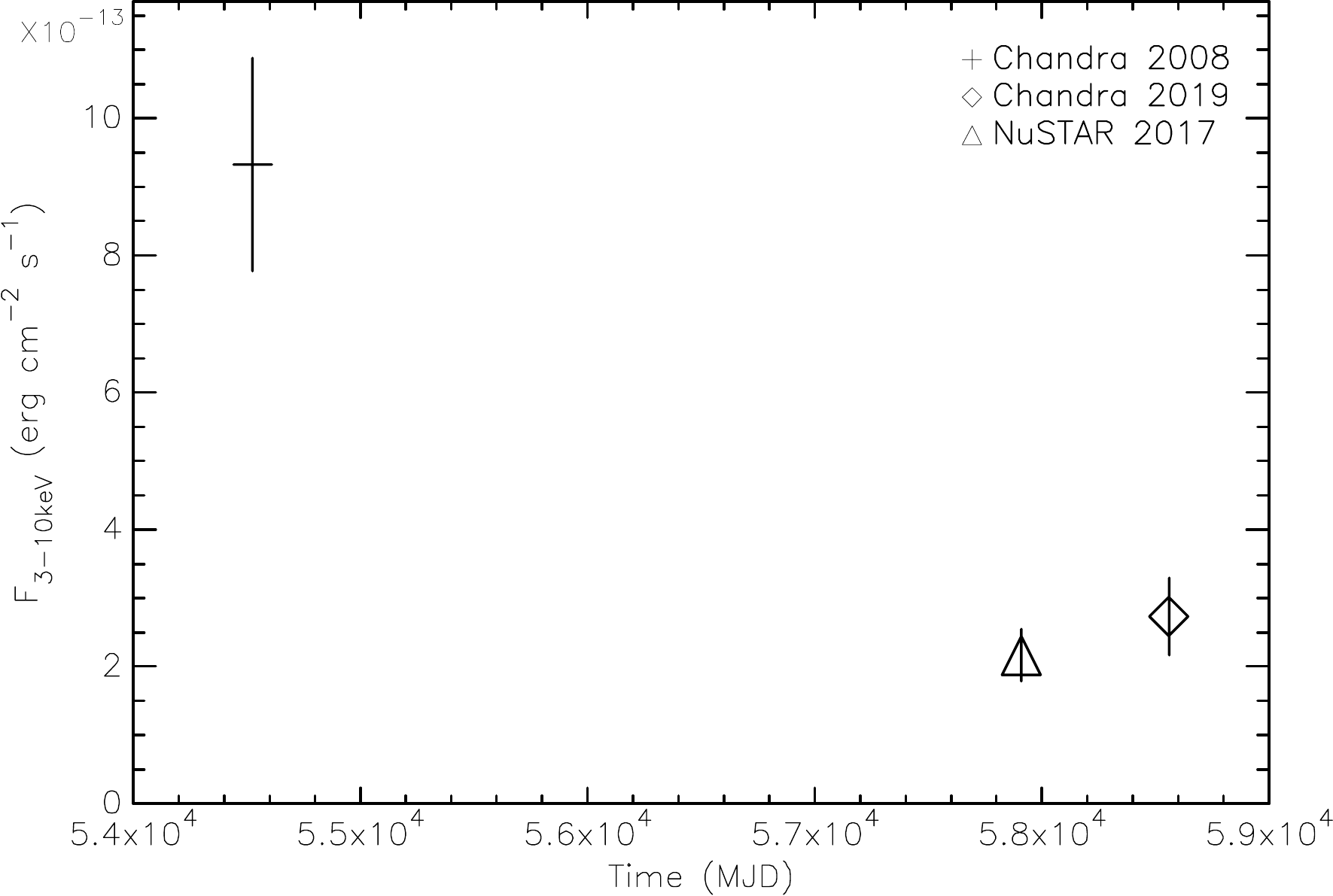}
  \caption{3--10~keV lightcurve of \xraysrc\ over the last decade from two \chandra\ observations (in 2008 and 2019) and one \nustar\ observation (in 2017). The source was brighter in 2008 than in 2017--2019 by a factor of $\sim4$. { The H.E.S.S. observations of this region as part of Galactic Plane Survey were carried out from 2004 to 2013, while the VERITAS archival data were collected from 2007 to 2015. Note that HAWC observed the source more recently in 2015--2017.}}
    \label{fig:flux}
  \end{center} 
\end{figure}

We extracted \nustar\ spectra of \xraysrc\ from a $r=30\asec$ region and generated the \nustar\ response matrix and ancillary response files using {\tt nuproducts}. Background spectra were extracted from a $r=60''$ source-free region on the same detector chip. The net count rate in the 3--20~keV, after combining FPMA and FPMB spectra, is 0.003 cts\,s$^{-1}$.  The 2008 \chandra\ spectra (ObsID 9081) are extracted from a $r_{\rm major/minor}=10/5''$ elliptical and a $r=10''$ circular region for the source (which is located at a large off-axis position)  and background, respectively with the net source count rate of  0.012\,cts\,s$^{-1}$. The 2019 (ObsID 22145)  \chandra\ spectra are extracted from a $r=2\asec$ circular region and $r=3-5\asec$ annular region for the source and background, respectively, yielding the 2--8~keV net count rate of 0.005\,cts\,s$^{-1}$. The response files are generated with the {\tt specextract} tool of {\tt CIAO }~4.11. We grouped the spectra to ensure at least  30 cts/bin for \nustar\ and 5 cts/bin for \chandra, and jointly fit the spectra with an absorbed power-law model, employing the {\tt lstat} statistic in {\tt XSPEC} v12.10.1. Employing other statistics such as {\tt cstat} or $\chi^2$ with {\tt gehrels} weight does not alter the results significantly. We find that the  absorbed power-law model  with $\Gamma_X= 1.6\pm0.3$ and $N_{\rm H} = (9\pm3)\times10^{22}$~cm$^{-2}$ describes the data well; the spectra are shown in  Figure~\ref{fig:spectra}. However, we find that the source flux as measured by \chandra\ varies by a factor of 4--5 over 9 years. The large $N_{\rm H}$ implies a large distance to the source, and the 2--8\,keV luminosity is estimated to be $L_X = (0.6-1.9)\times10^{33}$~erg\,s$^{-1}$ in the low-flux state for an assumed distance of 5--10\,kpc. 
Note that both \nustar\ and \chandra\ 2017--2019 observations yield X-ray (2-8 keV, absorbed) fluxes that are lower than the 2008 \chandra\ flux by a factor of 4--5 
(Figure~\ref{fig:flux}) at the $\simgt4\sigma$ significance level.
We also searched for an Fe line emission but did not find any significant emission.

For \nustar\ timing analysis, we extracted source photon events within a $r=30$\asec\ circle around the X-ray source. We then constructed 3--20~keV \nustar\ light curves and subtracted background light curves after the proper normalization. We found no significant modulation in the \nustar\ lightcurves. 
Furthermore, we found no evidence of aperiodic variability (i.e. red noise) in the power density spectra produced from the \nustar\ data. The 3--20 keV power density spectra are consistent with a flat white noise component, unlike accreting X-ray pulsars which often shows strong red noise components \citep{Lazzati1997}. 

\subsection{A potential IR counterpart of \xraysrc}

We searched IR-to-optical catalogs for a counterpart of the X-ray source CXO1928. There is a bright IR source (2MASS~J19281204$+$1747111) with
magnitudes of $J = 17.8$, $H=14.7$ and $K=13.6$, offset by $2.2$\asec\ from the \chandra\ position of \xraysrc. Although the offset is large compared to the statistical error of the \chandra\ position of 0.9\asec\ (1-$\sigma$), 
 the systematic errors can be as large as $\sim1.4$\asec\ (99\% CL)\footnote{\url{https://cxc.harvard.edu/cal/ASPECT/celmon/}}. Unfortunately, there is no X-ray source in the \chandra\ FOV that can be registered to its known IR or radio position for a boresight correction. Therefore, the association cannot be firmly established with the source positions. Alternatively, based on the surface density of nearby IR sources (which are  brighter than $K = 13.6$) in the 2MASS catalog, we estimated that the probability of chance coincidence between the \chandra\ and IR sources is 3\%; their association is only suggestive. Assuming an optical extinction $A_V>17$, as indicated by the large $N_{\rm H}$ measured from X-ray spectral fitting and using the relation in \citet{Guver2009}, we fit a stellar SED model to the existing IR flux data from Spitzer, UKIRT, and 2MASS which suggests the IR source is a massive star (likely O-type).
 

\section{Discussion} \label{sec:discussion}  

In this section, we consider three scenarios for the TeV emission from \hawcsrc: (1) a putative PWN of the radio pulsar \pulsar, (2) hadronic interactions in the unseen SNR shock  and (3) a TeV gamma-ray binary (TGB). Our investigation is largely based on the multi-wavelength SED including \chandra, \nustar, \fermi\ and HAWC data. In the first two cases, we assume that the variable X-ray source \xraysrc\ is not associated with the HAWC source, and therefore the X-ray SED is unconstrained. In the TGB case, we assume that both the X-ray source and bright IR counterpart candidate are associated with the HAWC source. We do not consider an extra-galactic origin, such as hard-TeV blazars  \citep{Magic2019} because the scenario seems implausible due to the lack of a radio galaxy counterpart \citep{Velzen2012} as well as no short-term ($\sim$ hours) variability from the X-ray source \citep{Pandey_2017}. 
\subsection{A putative PWN of the radio pulsar \pulsar} \label{sec:pulsar}

\pulsar\ is one of the leading counterpart candidates for the HAWC source due to its positional coincidence. However, no nebula has been detected around the pulsar in the radio band \citep{Cordes2006}. 
\chandra\ observations yielded no X-ray detection of the pulsar,
setting an upper limit of the unabsorbed flux in the 2--8\,keV band $F_{\rm X} < 7\times10^{-15}$~\eflux\ and $ < 3\times10^{-13}$~\eflux, assuming that the putative PWN is extended over $r=3$\asec\ and $r=4$\amin, respectively.  

In PWN models, it is believed that synchrotron radiation produces radio to X-ray photons, and inverse Compton upscattering of the synchrotron (self-Compton; SSC) and/or the external IR/CMB radiation fields produces the TeV emission. 
In order to bound some of the PWN parameters, we applied the leptonic model, {\tt InverseCompton+Synchrotron}, in the {\tt naima} software package \citep{Zabalza2015} to the multi-wavelength SED data. 
If we assume a compact PWN ($r \simlt 1$\amin), the very-high TeV-to-X-ray flux ratio of \hawcsrc\ ($F_{\rm TeV} / F_{\rm X} \simgt 100$) requires the PWN B-field far below the typical ISM value \citep[$B_{\rm PWN} \simlt 1\mu$G whereas $B_{\rm ISM} \sim 10$    $\mu$G; ][]{Crutcher2012}  and/or extremely high NIR and FIR  densities at  $U_\gamma \sim10^3-10^4$~eV\,cm$^{-3}$ for the typical PWN B-field range \citep[$B \sim 10-100\mu$G;][]{Martin2014}.  
The latter case is implausible since such a high radiation density in the IR/optical band is only observed in the Galactic Center \citep{Davidson1992}.  Alternatively, the PWN radius can be reduced to $R_{\rm PWN} \sim10^{-5}$~pc, as a result of SNR reverse shocks crushing the nebula \citep{Reynolds1984}, in order to amplify the SSC component to fit the TeV spectra. 
However, the shock compression amplifies the magnetic field strength, and thus the synchrotron emission should be detectable in the radio and X-ray bands \citep{Gelfand2007}. 

\begin{figure}
\begin{center} 
\includegraphics[width=1.0\linewidth]{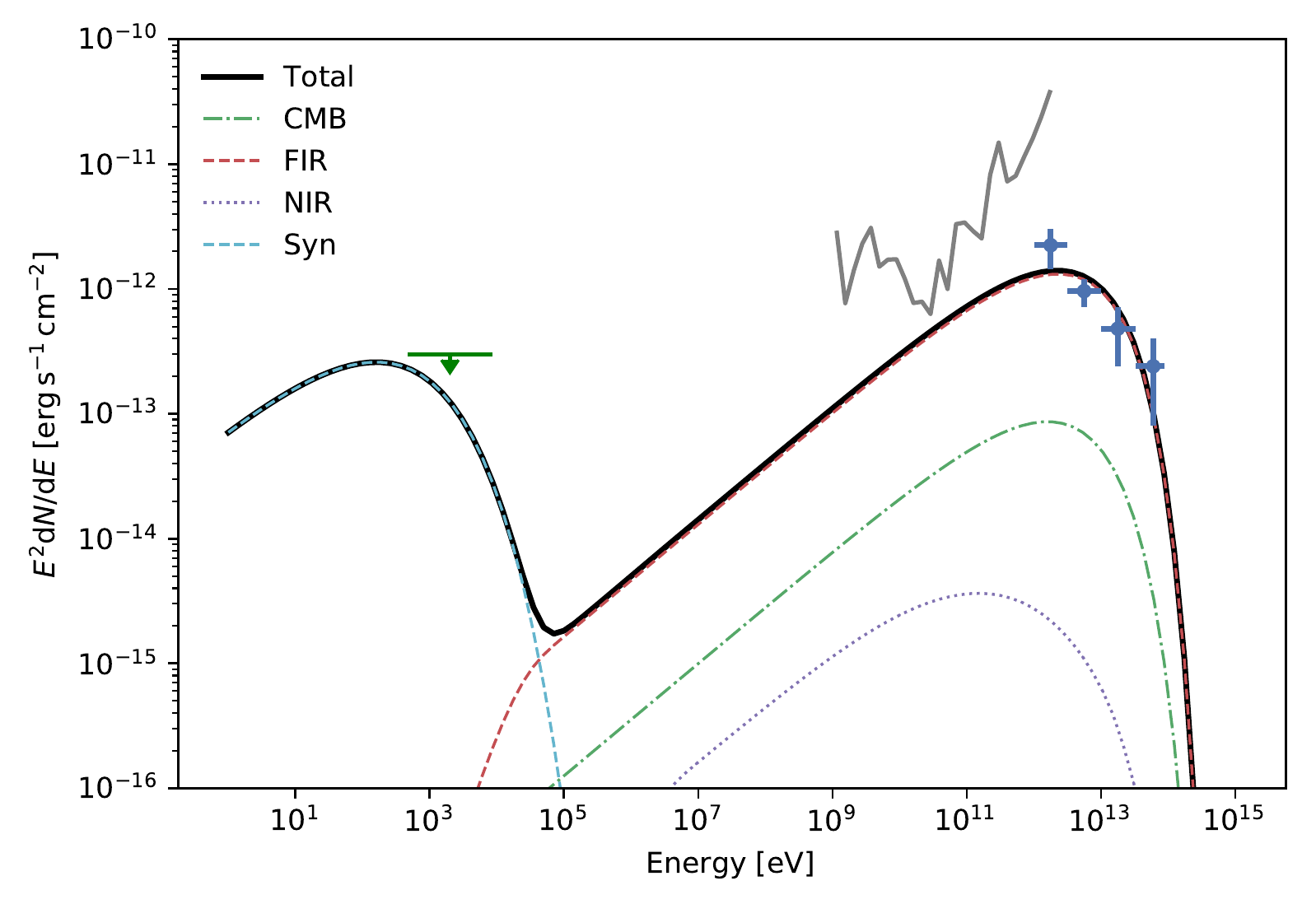}
  \caption{A SED plot for the PWN scenario with $R_{\rm PWN}=6$~pc. The \fermi-LAT flux upper limits and HAWC TeV flux data are plotted as the grey line and blue points, respectively. The X-ray flux upper limits (green arrow) were determined from a $r=4$\amin\ circle around the radio pulsar position using the \chandra\ ACIS data. The radiation densities in the NIR and FIR bands are set to the values of 1 eV\,cm$^{-3}$ and  10 eV\,cm$^{-3}$. We adopted a cutoff power-law model for the electron energy spectrum with $\alpha_e  = 2.1$ and $E_{\rm cut}$ = 30~TeV and a PWN magnetic field strength of $B = 5 \mu$G as a representative case. }
 \label{fig:sed_pulsar}
 \end{center} 

\end{figure}

We found that the pure leptonic case is marginally plausible only when we assume a large PWN size of $r \sim 4$\amin\ or $r\sim6$~pc at the   pulsar distance of 5.8~kpc \citep{Nice2013}, thus relaxing the X-ray flux upper limit to $<  3\times10^{-13}$~\eflux. According to \citet{Bamba2010} who studied the X-ray PWN size variation with the spin-down age, the 83-kyr old pulsar may well be extended beyond $r\sim6$~pc. For example, as shown in Figure~\ref{fig:sed_pulsar}, the SED data from X-ray to TeV bands can be fit with $B_{\rm PWN} \sim 5 \mu$G and elevated IR radiation densities ($U_{\rm NIR} = 1$~eV\,cm$^{-3}$ and $U_{\rm FIR} = 10$~eV\,cm$^{-3}$). \citet{Bamba2010} argued that such a extended, faint PWN can have its magnetic field strength decayed to below the typical ISM $B$-field. 
Also,  more recent observations of the region using \textit{WISE} and \textit{Gaia} data found five star clusters (Cmg 495, Cmg 497, Cmg 498, Cmg 499, and Liu \& Pang catalog ID 1262) within $r\sim8$\amin\ from the pulsar position, some of which could be embedded in molecular clouds \citep{Carmargo2015, Liu2019}. These star clusters can contribute to enhancing ICS emission to the TeV flux level observed by HAWC. Therefore, we conclude that a diluted, unseen PWN associated with the radio pulsar can account for the TeV emission. A deeper X-ray survey around the pulsar may uncover diffuse X-ray emission like other faint X-ray PWNe detected by \citet{Bamba2010}.



\begin{figure}
\begin{center} 
\includegraphics[width=1.0\linewidth]{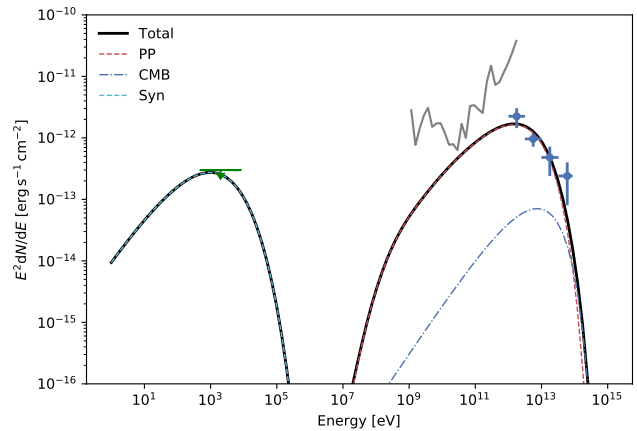}
\caption{A SED plot in the hadronic scenario, using the hadronic {\tt PionDecay} and leptonic models in the {\tt naima} package. The \fermi-LAT flux upper limits and HAWC TeV flux data are plotted as the grey line and blue points, respectively. For the leptonic model, we assumed only the CMB as a source of seed photons for the ICS component and a magnetic field strength of $B=5~\mu$G. The X-ray flux upper limits (green arrow) were determined from a $r=4$\amin\ circle around the radio pulsar position using the \chandra\ ACIS data. Parameters are $\alpha_p = 1.4$, $E_{\rm cut}$ = 40 TeV and $n_{\rm H} = 130 $ cm$^{-3}$ (i.e., the mean hydrogen density of the molecular cloud G52.9$+$0.1). We adopted the same parameters for the electron and proton energy spectra except for the normalization factors. The total energy of the relativistic protons is $W_{\rm p} = 4.8\times10^{47}$ erg (1--10 TeV) or $ 2.1\times10^{48}$ erg (0.01--100 TeV). The total energy of the relativistic electrons should be lower than  $W_{\rm e} = 7.1\times10^{45}$ erg (1--10 TeV) or $ 2.1\times10^{46}$ erg (0.01--100 TeV).  
}
\label{fig:sed_hadronic}  
\end{center} 
\end{figure}

\subsection{Hadronic interactions} \label{sec:hadronic} 

Alternatively, the TeV emission could originate primarily from hadronic interactions as a result of collisions between relativistic protons and the ISM or nearby molecular clouds. Pion decays from proton-proton collisions are efficient TeV emitters, whereas the ICS component from $\simgt 100$-TeV  electrons is  suppressed  at $E_\gamma \simgt 10$~TeV  due to the Klein-Nishina effect \citep{Rieger2013}. 
The molecular cloud G52.98$+$0.18 in the HAWC source region can serve as a target for hadronic interactions \citep{Rice2016}. From the molecular cloud's measured mass, angular size, and distance of 4.49$\times10^{5}$ M$_\sun$, 0.22$^{\circ}$, and 9.56 kpc, respectively, we estimated that the average hydrogen density is 130 cm$^{-3}$. 
There is no radio or X-ray SNR within $r\sim30$\amin\ from the HAWC source; however, soft X-ray emission from the putative SNR may be strongly absorbed. We assumed a putative SNR with the shell radius derived from the Sedov solution of $r_s \propto (E t^2_0/n_0)^{1/5}$ where $E = 10^{51}$~erg (the total SN energy released), $t_0 = 83$~kyrs (the spin-down age of the radio pulsar) and $n_0$ (the mean number density of the molecular cloud). We estimated that the SNR shell radius should be $\sim11$~pc or $\sim4$\arcmin\ assuming that the source is located at the distance of the molecular cloud at 10~kpc \citep{Rice2016}. Alternatively, if we adopt the distance to the pulsar \citep[5.8~kpc;][]{Nice2013}, the angular radius of the SNR  should be $\sim7$\amin. 


To explore the hadronic scenario, we applied a combination of the leptonic and hadronic models in the {\tt naima} package to the multi-wavelength SED data. 
We assumed that the particle energy spectrum follows a power-law with an exponential cutoff ($N(E) \propto E^{-\alpha}e^{-E/E_{\rm cut}}$) because a single power-law model does not fit the SED. First, we fit the {\tt PionDecay} model only in the gamma-ray band. A hard proton spectral index ($\alpha_p = 1.4$) and an exponential cutoff at $E_{\rm cut} = 40$~TeV are required to give a SED model consistent with the \fermi\ GeV upper limit and HAWC TeV spectra. The total energy of relativistic protons (0.01--100~TeV) is $W_p = 2.1\times 10^{48}$~erg.  This corresponds to a small fraction ($<0.1$\%) of the typical supernova energy ($\sim10^{51}$~erg).  

Since the {\tt naima} package does not track the by-product leptons from the {\tt PionDecay} model, we added a separate leptonic model to constrain the secondary electron population so that their synchrotron emission does not over-predict the radio and X-ray flux upper limits. Note that the ICS component from the electron population needs to have a small contribution so as to not overshoot the TeV fluxes.  Assuming the same energy spectrum for electrons and protons (i.e., $N(E) \propto E^{-\alpha}e^{-E/E_{\rm cut}}$ with $\alpha = 1.4$ and $E_{\rm cut} = 40$~TeV) and a typical ISM magnetic field strength \citep[$B_{\rm ISM} = 5$ $\mu$G;][]{Crutcher2012}, we found that the total energy of relativistic electrons ($N_e$) should be less than $W_e = 2.1\times10^{46}$~erg (0.01--100~TeV) so as not to exceed the X-ray flux upper limit from a $r=4$\amin region around the radio pulsar  (Figure~\ref{fig:sed_hadronic}). 
If the putative SNR is located at the molecular cloud distance ($\sim10$~kpc), its angular size is estimated to  $r\sim7$\amin, which is larger than the FOV of the \chandra\ ACIS observations. Assuming that diffuse X-ray emission associated with the putative SNR is spatially uniform, the X-ray flux upper limit is higher by a factor of $\sim3$ thus enhances the electron energy to $W_e = 6.3\times10^{46}$~erg.  
However, the small ratio of $W_e/W_p \sim 0.01$--0.03 is still difficult to reconcile with the p-p collision case since the total kinetic energy of secondary electrons (i.e. byproducts of charged pion decays) should be $\sim1/3$ of the total radiation energy of $\pi^0$ gamma rays \citep{Coerver2019}. In order to yield the ratio $W_e / W_p$ comparable to $\sim 1/3$, it requires a lower ambient $B$-field, which seems implausible within a molecular cloud where the B-field should be amplified. Only if the extent of relativistic proton population is larger than $r\sim7$\arcmin\ (e.g., the SNR may be older than the spin-down age), the resultant synchrotron emission may be consistent with the X-ray flux limits while $B_{\rm ISM} \sim 5 \mu$G.   


It is therefore possible that relativistic protons in a diluted, undetected SNR, extending over a $r \simgt 4$\arcmin\ region, produce TeV gamma rays via collisions with the molecular clouds. Thus, diffuse X-ray or radio emission associated with the  HAWC source could be too faint to be detected \citep{Butt2008}.    
Prior to observations by the Cherenkov Telescope Array (CTA) which may resolve the TeV emission \citep{CTA2019}, it is essential to survey a  larger region around the HAWC source with X-ray telescopes. Some TeV sources are associated with extended diffuse X-ray sources, and even non-detection of an X-ray counterpart can be useful for inferring the source type. 
\xmm, with its large FOV, is best suited to search for other X-ray counterpart candidates in a larger region around the HAWC source. 



\subsection{A new TeV gamma-ray binary? } 

If the variable X-ray source \xraysrc\ is associated with the HAWC source, it may belong to a rare class of TeV gamma-ray binaries \citep{Dubus2013}.
These systems are likely composed of a neutron star (NS)  orbiting
around a massive O/B star. The exact mechanism responsible for the non-thermal emission is still unknown. However, a possible scenario is that $e^\pm$ pairs from the pulsar wind are accelerated at the shock produced by the interaction between the pulsar and the stellar wind. The resultant high-energy electron population emits synchrotron and ICS radiation which accounts for the observed X-ray and gamma-ray emission, respectively~\citep{Tavani1997}. 

\begin{figure}
\begin{center} 
  \includegraphics[width=1.0\linewidth]{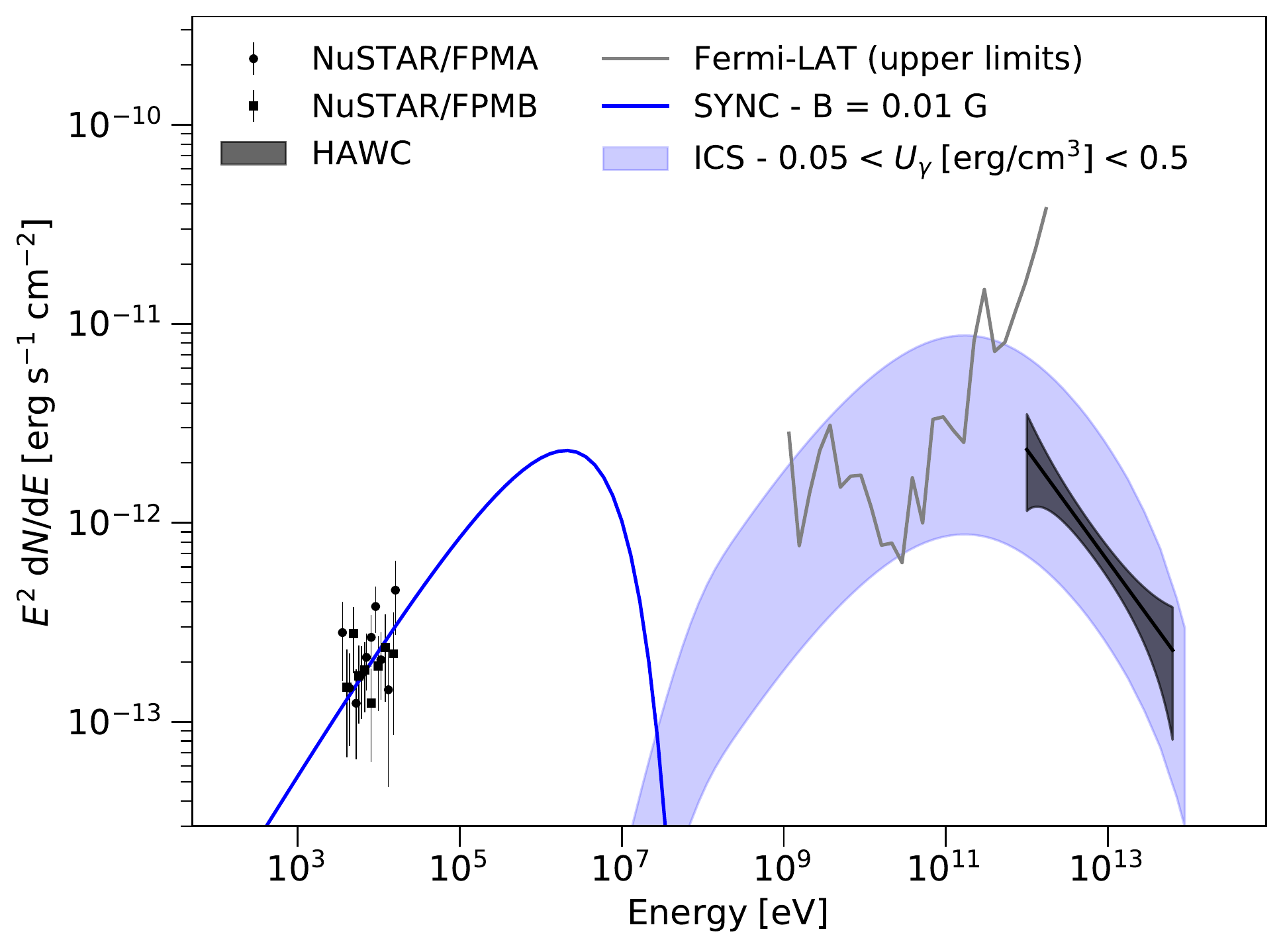}
  
  \includegraphics[width=1.0\linewidth]{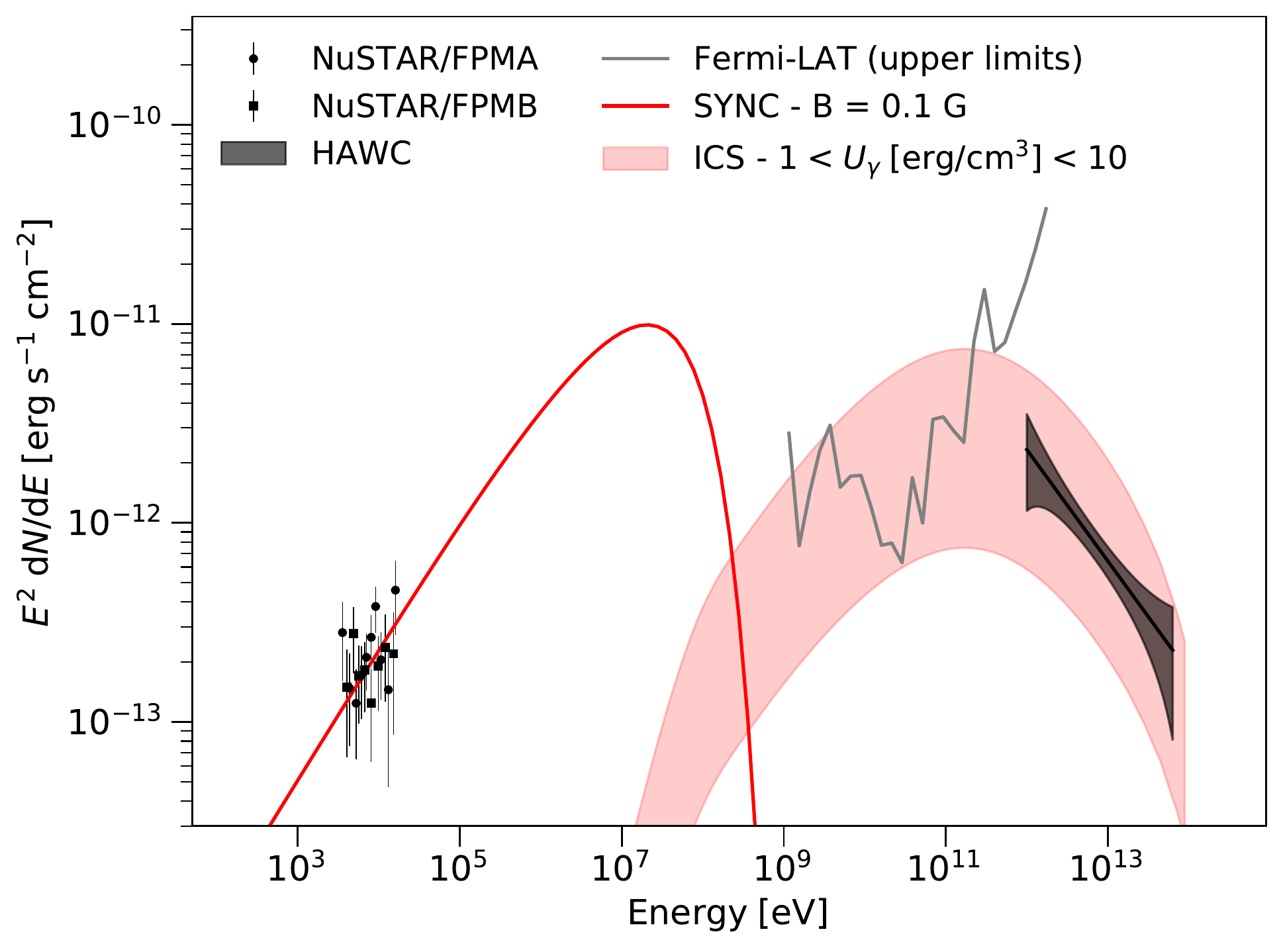}
    \caption{A TGB model fit to the X-ray and gamma-ray SED data with $B=0.01$~G (top panel) and $B=0.1$~G (bottom panel). The \fermi/LAT 5-$\sigma$ upper limits are indicated by dotted lines. The colored lines and bands represent the synchrotron and ICS components of the TGB SED model, respectively. }  
    \label{fig:sed_tgb}
\end{center} 
\end{figure}

The X-ray spectral and timing signatures of \xraysrc\ -- a single power-law spectrum with $\Gamma_X \sim 1.6$, long-term time variability and X-ray luminosity  -- are consistent with those of other TGBs with massive (O or B) companion stars~\citep{Dubus2013}; such variability can be explained as due to varying $B$ or bulk Doppler boost as the system geometry changes with orbital phases \citep[e.g.,][]{ar17}. Unlike accreting X-ray binaries, the lack of X-ray aperiodic variability shorter than a day is consistent with the TGB scenario~\citep{Mori2017}. 
In addition, both the TeV gamma-ray photon index ($\Gamma_{\rm{TeV}} = 2.6$) and the TeV to X-ray flux ratio ($F_{\rm{TeV}}/F_X = 0.4-2$) are in the typical parameter range for other TeV gamma-ray binaries \citep{Dubus2013}.

In order to further probe the TGB hypothesis, we compared the \nustar\ and HAWC SED data to a generic, analytical model based on the NS assumption \citep[see more details in][] {Archer2019}. In this model the energy spectrum of the high-energy electron population is described by a power law with an exponential cutoff. While the normalization and the slope of the power law were obtained by fitting the \nustar\ data, the cutoff energy ($E_{\rm cut}$) was set to 100 TeV, which is assumed to be a minimum value to describe the HAWC observations. The typical $B$-field strength within this scenario is 0.01--1 G~\citep{Archer2019}. In Figure~\ref{fig:sed_tgb}, we show the SED comparison in two cases in which $B=0.01$ G (top) and $B=0.1$ G (bottom). For values higher than that, it was found that the expected flux in the GeV band is too large and could not accommodate the lack of detection by Fermi-LAT. 

From the high-energy electron spectrum obtained by fitting the \nustar\ data, we calculated the expected flux of gamma rays produced by ICS. The ICS photon field in this scenario is given by thermal UV photons from the O/B companion star, which we assumed to be at the typical temperature of $3\times10^4$ K. The photon density ($U_{\gamma}$) varies substantially with the distance of the shock from the companion star and therefore is strongly dependent on the geometry of the system which is unknown. 
{ Besides that, $\gamma-\gamma$ absorption within the system, which is also geometry dependent, may also affect the observed gamma-ray flux. Due to these limitations, it is only possible to loosely constrain the SED in the gamma-ray band.}
In Figure~\ref{fig:sed_tgb}, we show filled bands for the ICS spectra for a given range of $U_{\gamma}$ which can be reasonably expected for TGBs and would also be consistent with the HAWC data. Although the \nustar\ and the HAWC data were taken at different epochs and TGBs are very variable sources, the SED comparison shows that the  data are consistent with the expectation from a TGB scenario. However, further broadband studies are required to confirm the TGB scenario, and particularly, detection of TeV variability would be a smoking gun. { The non-detection of the HAWC source by VERITAS may indicate  source variability similar to the TeV gamma-ray binary HESS J0632$+$057 which was initially not detected by VERITAS \citep{Acciari2009}.} If confirmed, \hawcsrc\ may be a unique binary system emitting gamma rays up to $\sim100$~TeV since no other known TGBs have been detected by HAWC above $E\sim10$~TeV \citep{Rho2017}. 


\section{Summary} \label{sec:summary} 

\begin{itemize} 

\item \hawcsrc\ is one of the Galactic TeV sources detected by HAWC up to $\sim56$~TeV and { later confirmed by H.E.S.S. assuming an extended source. The non-detecion by VERITAS also suggests that the TeV emission may be extended or variable. }
There is no SNR, PWN or \fermi\ source coinciding with the HAWC source position.  

\item The 83-kyr-old radio pulsar \pulsar\ can account for the TeV emission in the pure leptonic scenario, only if its putative PWN is extended beyond $r\sim4$\amin. An alternative scenario could involve an unseen SNR whose shock produces the TeV emission via relativistic protons colliding with a nearby molecular cloud. The estimated proton energy ($W_p = 4.7\times10^{47}$~erg) is reasonable as only a small fraction ($\sim0.1$\%) of the supernova energy is required to power the TeV emission. In the hadronic scenario, diffuse X-ray emission from secondary electron synchrotron radiation over $r \simgt 7$\arcmin\ should be present in the region. 

\item \nustar\ and \chandra\ detected a bright X-ray source \xraysrc\ which overlaps  with the HAWC source position. The non-thermal X-ray spectra, long-term X-ray flux variation, the lack of aperiodic variability on a time scale shorter than a day and a potential association with a bright IR source suggests the HAWC source may be a new TGB. However, it needs to be confirmed by detecting variability in the TeV band. 
{ Even if the HAWC source is extended, as suggested by the H.E.S.S and VERITAS observations, a TGB may be still present in the region. For example, in follow-up observations of the extended TeV source TeV J2032$+$4130 \citep{Aliu2014b},  VERITAS recently detected PSR J2032$+$4127/MT91 213,  a TeV gamma-ray  binary system with a 50~year orbital period  \citep{Abeysekara2018b}, spatially coincident with the extended source. } 


\item The HAWC source, given its location in a complex region with star clusters and molecular clouds, may be composed of several TeV sources such as a faint (undetected) nebula of the radio pulsar and diffuse TeV emission from hadronic interactions and may be spatially resolved by the near-future CTA observatory. Until then, a large/deep X-ray survey around the HAWC source may provide clues of the source type. \xmm, given its large FOV, is best suited to search for other X-ray counterpart candidates in a larger region around the HAWC source. 

\end{itemize} 
\acknowledgments
 
This work made use of data from the \nustar\ mission, a project led by the California Institute of Technology,
managed by the Jet Propulsion Laboratory, and funded by the
National Aeronautics and Space Administration. We thank the \nustar\  Operations, Software and Calibration teams for support with the execution and analysis of these observations.
This research has made use of the \nustar\  Data Analysis
Software (NuSTARDAS) jointly developed by the ASI Science
Data Center (ASDC, Italy) and the California Institute of
Technology (USA). 
This  research  is  supported  by  grants  from the  U.S.  Department  of  Energy  Office  of  Science, the U.S. National Science Foundation and the Smithsonian Institution, and by NSERC in Canada.  We acknowledge the excellent work of the technical support staff at the Fred Lawrence Whipple Observatory and at the collaborating institutions  in  the  construction  and  operation of the VERITAS instrument.  H.A. acknowledges support from Basic Science Research Program through
the National Research Foundation of Korea (NRF)
funded by the Ministry of Science, ICT \& Future Planning (NRF-2017R1C1B2004566).  J.H. acknowledges support from an appointment to the NASA Postdoctoral Program at the Goddard Space Flight Center, administered by the USRA through a contract with NASA. O.K. was supported by the National Aeronautics and Space Administration through the award 80NSSC19K0576. { Q.F. and R.M. acknowledges support from NSF Grant PHY-1806554. }

%



\vspace{0.4cm} 
{\textit Software:} HEASoft Version 6.25 \citep{heasoft}, FTools Version 6.25 \citep{heasoft}, naima Version 0.9.0 \citep{Zabalza2015}



\bibliography{2hwc1928} 



\end{document}